\documentclass[showpacs,preprintnumbers,twocolumn]{revtex4}

\usepackage[centertags]{amsmath}
\usepackage{amsfonts}
\usepackage{amssymb}
\usepackage{amsthm}
\usepackage{newlfont}
\usepackage{graphicx}
\input{epsfx}

\begin{document}

\title{Efficient bounds on quantum communication
rates {\it via} their reduced variants}
\author{Marcin L. Nowakowski and Pawel Horodecki\footnote{Electronic address: pawel@mif.pg.gda.pl}}
\affiliation{Faculty of Applied Physics and Mathematics, ~Gdansk
University of Technology, 80-952 Gdansk, Poland}
\affiliation{National Quantum Information Centre of Gda´nsk,
Andersa 27, 81-824 Sopot, Poland}


\begin{abstract}
We investigate one-way communication scenarios where Bob manipulating on his parts can transfer some sub-system to the environment. We define reduced versions of quantum communication rates and further, prove new upper bounds on one-way quantum secret key, distillable entanglement and quantum channel capacity by means of their reduced versions. It is shown that in some cases they drastically improve their estimation.
\end{abstract}

\maketitle Recently years have seen enormous advances in quantum
information theory proving it has been well established as a basis
for a concept of quantum computation and communication. Much work
\cite{BennettDiVincenzo, BennettSmolin, Barnum3, Barnum4,
DevetakW1, DevetakW2, DevetakW3} has been performed to understand
how to operate on quantum states and distill entanglement enabling
quantum data processing or establish quantum secure communication
between two or more parties. One of the central problems of
quantum communication field is to estimate efficiency of
communication protocols establishing secure communication between
involved parties or distilling quantum entanglement \cite{Renner,
KHPH, KHPH2, DevetakW1, DevetakW2, DevetakW3, Smith}. Most simple
communication scenarios are those that do not use classical side
channel or use it only in one-way setup. The challenge for the
present theory is to determine good bounds on such quantities like
the secret key rate or quantum channel capacity and distillable
entanglement of a quantum state, that allow to estimate the
communication capabilities. In this paper we provide efficient
upper bounds avoiding massive overestimation of communication
rates. We are inspired by classical information and entanglement
measures theory where so-called reduced quantities have been used
\cite{Renner, KHPH, DiVincenzo}. Herewith we consider two pairs of
quantities: private capacity $\mathcal{P}$, quantum one-way secret
key $K_{\rightarrow}$ and one-way quantum channel capacity
$\mathcal{Q}_{\rightarrow}$, one-way distillable entanglement
$D_{\rightarrow}$ providing new efficient upper bounds. We prove
that in some cases the bounds explicitly show that the
corresponding quantity is relatively small if compared to sender
and receiver systems. The main method is again the fact that all
the above quantities vanish on some classes of systems. Moreover,
we introduce 'defect' parameters $\Delta$ for the reduced
quantities resulting from possible transfer of sub-systems on
receivers' side which are (sub)additive and hence, can be
exploited in case of composite systems and regularization.

\textit{\bf Reduced one-way secret key.} A secret key is a quantum
resource allowing two parties Alice and Bob private communication
over a public channel. In an ideal scenario they generate a pair
of maximally correlated classical secure bit-strings such that Eve
representing the adversary in the communication is not able to
receive any sensible information from further communication
between Alice and Bob. In this section we will elaborate on
generation of a one-way secret key from a tripartite quantum state
shared by the parties with Eve that means Alice and Bob can use
only protocols consisting of local operations and one-way public
communication. We propose a new reduced measure of the one-way
secret key that simplify in many cases analysis of one-way
security of quantum states.

To derive new observations about one-way quantum secret key we utilize in this section fundamental information notions engaging entropy \cite{Entropy} and quantum mutual information \cite{MInformation} which play a vital role in quantum information theory.
We state a new result about the upper bound on the Holevo function \cite{HolevoFunction} $\chi(\cdot)$:


\textbf{Observation 1.}\label{reducedholevo2} \textit{For any
ensemble of density matrices $\mathfrak{A}=\{\lambda_{i},
\rho^{i}_{BB'}\}$ with average density matrix
$\rho_{BB'}=\sum_{i}\lambda_{i}\rho^{i}_{BB'}$ there holds:
\begin{equation}\label{reducedholevo}
\chi(\rho_{BB'}) \leq \chi(\rho_{B}) + 2S(\rho_{B'})
\end{equation}}
\textit{Proof. }
Basing on subadditivity and concavity of quantum entropy we can easily show that:
\begin{eqnarray*}\label{LHS1}
&&|S(\rho_{BB'})-\sum_{i}p_{i}S(\rho^{i}_{BB'})-S(\rho_{B})+\sum_{i}p_{i}S(\rho^{i}_{B})| \leq \\
&\leq&|S(\rho_{BB'})-S(\rho_{B})|+ |\sum_{i}p_{i}S(\rho^{i}_{BB'})-\sum_{i}p_{i}S(\rho^{i}_{B})\\
&\leq&S(\rho_{B'})+\sum_{i}p_{i}S(\rho^{i}_{B'})\leq 2S(\rho_{B'})
\end{eqnarray*}
which completes the proof. $\Box$

One can use \cite{DevetakW1, DevetakW2} a general tripartite pure
state $\rho_{ABE}$ to generate a secret key between Alice and Bob.
Alice engages a particular strategy to perform a quantum
measurement (POVM) described by $Q=(Q_{x})_{x \in \cal X}$ which
leads to: $\widetilde{\rho}_{ABE}=\sum_{x}|x\rangle\langle x|_{A}
\otimes Tr_{A}(\rho_{ABE}(Q_{x})\otimes I_{BE})$. Therefore,
starting from many copies of $\rho_{ABE}$ we obtain many copies of
cqq-states $\widetilde{\rho}_{ABE}$ and we restate the theorem
defining one-way secret key $K_{\rightarrow}$:

\textbf{Theorem 1.}\cite{DevetakW1}
\textit{For every state $\rho_{ABE}$,
$K_{\rightarrow}(\rho) = \lim_{n\rightarrow\infty}\frac{K_{\rightarrow} ^{(1)}(\rho^{\otimes n})}{n}$,
with
$K_{\rightarrow}^{(1)}(\rho)=\max_{Q,T|X} I(X:B|T) - I (X:E|T)$
where the maximization is over all POVMs $Q=(Q_{x})_{x \in \cal X}$ and channels R
such that $T=R(X)$ and the information quantities refer to the state:
$\omega_{TABE}=\sum_{t,x} R(t|x)P(x)
|t\rangle\langle t|_{T}\otimes |x\rangle\langle x|_{A} \otimes
Tr_{A}(\rho_{ABE}(Q_{x})\otimes I_{BE}).$
The range of the measurement Q and the random variable T may be assumed to be bounded as follows:
 $|T|\leq d^{2}_{A}$ and $|\cal X|\leq d^{2}_{A}$ where T can be taken a (deterministic) function
 of $\cal X$.
}

Following we define a modified version of the one-way secret key rate $K_{\rightarrow}$ basing on the
results of \cite{Renner,KHPH} for reduced intrinsic information and reduced entanglement
measure.

\textbf{Definition 1.} \textit{For the one-way secret key rate
$K_{\rightarrow}^{(1)}(\rho_{AB})$ of a bipartite state
$\rho_{AB}\in B(\cal{H}_{A}\otimes \cal{H}_{B})$ shared between
Alice and Bob the reduced one-way secret key rate
$K_{\rightarrow}^{(1)}\downarrow(\rho_{AB})$ is defined as:
\begin{equation}
K_{\rightarrow}^{(1)}\downarrow(\rho_{AB})=\inf_{\cal{U}}[K_{\rightarrow}^{(1)}(\cal{U}(\rho_{AB}))+\Delta_{K_{\rightarrow}}
]
\end{equation}
where $\cal{U}$ denotes unitary operations on Bob's system with a
possible transfer of subsystems from Bob to Eve, i.e.
$\cal{U}(\rho_{AB})=Tr_{B'}(I\otimes \cal{U})\rho_{ABB'}$.
$\Delta_{K_{\rightarrow}} =4 S (\rho_{B'})$ denotes the defect
parameter related to increase of entropy produced by the transfer
of B'-subsystem from Bob's side to Eve.}

The reduced one-way secret key rate is an upper bound on $K_{\rightarrow}$ which we prove now
for every cqq-state $\rho$:

\textbf{Theorem 2.} \textit{For every cqq-state $\rho_{ABE}$ there
holds:
\begin{equation}
K_{\rightarrow}(\rho)=\lim_{n\rightarrow\infty} \frac
{K_{\rightarrow}^{(1)}(\rho^{\otimes n})}{n} \leq
K_{\rightarrow}\downarrow(\rho)
\end{equation}
where $K_{\rightarrow}\downarrow(\rho)=\lim_{n\rightarrow\infty}
\frac { K_{\rightarrow}^{(1)}\downarrow(\rho^{\otimes
n})}{N}$.Particularly, for identity operation $\cal{U}=id$ on
Bob's side one obtains: $K_{\rightarrow}(\rho_{ABB'}) \leq
K_{\rightarrow}(\rho_{AB})+ 4S(\rho_{B'})$.}

To prove this theorem one can start showing how the formula behaves for one-copy secret key:

\textbf{Lemma 2.}
\textit{
For every cqq-state $\rho_{ABE}$ there holds:
\begin{equation}\label{lemma2}
K_{\rightarrow}^{(1)}(\rho)
\leq K_{\rightarrow}^{(1)}\downarrow(\rho)
\end{equation}}
\begin{proof}
Since
\[
 \left\lbrace
           \begin{array}{l}
  I(A:B|C)=S(AC)+S(BC)-S(ABC)-S(C)\\
  I(A:E|C)=S(AC)+S(EC)-S(AEC)-S(C)\\
\end{array}
         \right.
\]
then:
\[
K_{\rightarrow}^{(1)}(\rho)=\max_{Q,C|A}[S(BC)-S(ABC)-S(EC)+S(AEC)]
\]

To prove the thesis of this lemma it suffices to show that:
\begin{equation}\label{key1}
K_{\rightarrow}^{(1)}(\rho_{A(BB')E})\leq K_{\rightarrow}^{(1)}(\rho_{AB(B'E)})+4S(B')
\end{equation}
due to the fact that in case of application of $\cal{U}$ without discarding subsystem $B'$ one
obtains equality. We denote by $\rho_{AB(B'E)}$ transition of $B'$-subsystem to the environment.
For (\ref{key1}) we can omit maximization that is
performed on both side of the inequality representing an application of a chosen 1-LOCC protocol distilling
a secret key that invokes:
\begin{eqnarray*}
&&S(BB'C)-S(ABB'C)-S(EC)+S(AEC) \leq \\
&&S(BC)-S(ABC)-S(B'EC)+S(AB'EC)+ 4S(B')
\end{eqnarray*}
It is easy to note that application of unitary operations on Bob's side do not change the inequality mainly
due to property of unitary invariancy of the von Neumann entropy.
To simplify the proof one can decompose this inequality into following two inequalities:
\begin{equation}\label{key2}
 \left\lbrace
           \begin{array}{l}
  S(BB'C)-S(ABB'C)\leq S(BC)-S(ABC) + 2S(B')\\
  S(B'EC)-S(AB'EC)\leq S(EC)-S(AEC) + 2S(B')\\
\end{array}
         \right.
\end{equation}
or equivalently considering the assumption that the initial state is of cqq-type and 'A' represents
classical distribution we can rewrite the first inequality into the form:
\begin{eqnarray*}
&&S(\sum_{i}p_{i}\rho_{i}^{BB'})-H(p_{i})-\sum_{i}p_{i}S(\rho_{i}^{BB'})-S(\sum_{i}p_{i}\rho_{i}^{B})\\
&&+H(p_{i})+\sum_{i}p_{i}S(\rho_{i}^{B})\leq 2S(B')
\end{eqnarray*}
and similarly for the second inequality which gives in result a more compact structure:
\[
 \left\lbrace
           \begin{array}{l}
\chi(\sum_{i}p_{i}\rho_{i}^{BB'C})-\chi(\sum_{i}p_{i}\rho_{i}^{BC})\leq 2S(B') \\
\chi(\sum_{i}p_{i}\rho_{i}^{B'EC})-\chi(\sum_{i}p_{i}\rho_{i}^{EC})\leq 2S(B') \\
\end{array}
         \right.
\]
However, the above was proved in Lemma 1 that completes the proof.
\end{proof}

Finally, we will extend this result in the asymptotic regime
proving Theorem 2.

\begin{proof} To prove Theorem 2 it suffices to notice that (\ref{lemma2})
holds under 1-LOCC and an arbitrary chosen $\mathcal{U}$ for any
$\rho_{n}=\rho^{\otimes n}$. Moreover, existence of the defect
parameter $\Delta_{K_{\rightarrow}}$ enables regularization of the
reduced one-way secret rate since in the asymptotic regime after
application of unitary operations on Bob side one can apply
subadditivity of entropy to estimate entropy of the transferred B'
part which implies $K_{\rightarrow}(\rho_{ABB'}) \leq
K_{\rightarrow}(\rho_{AB})+ 4S(\rho_{B'})$. \end{proof}

It is interesting that our results reflect E-nonlockability of the
secret key rate \cite{Ekert} which means that the rate cannot be
locked with information on Eve's side.

Monogamy of entanglement has been used to prove that for some
region quantum depolarizing channel has zero capacity even if does
not destroy entanglement \cite{Bruss} which is a particular
application of symmetric extendibility of states to evaluation of
the quantum channel capacity. The following examples will show
application of the concept:

\textit{Example 1.} As an example of application of Theorem 2 we
present a state which after discarding a small B' part on Bob's
side becomes a symmetric extendible state \cite{MNPH}. This
example is especially important since the presented state does not
possess \cite{MNPH2} any symmetric extendible component in its
decomposition for symmetric and
 non-symmetric parts, thus, one cannot use the method
\cite{Lutk3} to find an upper bound on $K_{\rightarrow}$ by means of linear optimization.
Let us consider a bipartite quantum state shared between Alice and Bob on the Hilbert
space $\cal{H}_{A}\otimes\cal{H}_{B}\cong \cal{C}^{d+2}\otimes\cal{C}^{d+2}$:

\begin{equation}
\rho_{AB}=\frac{1}{2} \left[
            \begin{array}{cccc}
              \Upsilon_{AB} & 0 & 0 & \cal A\\
              0 & 0 & 0 & 0\\
              0 & 0 & 0 & 0\\
              \cal A^{\dagger} & 0 & 0 & \Upsilon_{AB}\\
            \end{array}
          \right]
\end{equation}
where $\cal A$ is an arbitrary chosen operator so that $\rho_{AB}$
represents a correct quantum state. This matrix is represented in
the computational basis $|00\rangle, |01\rangle, |10\rangle,
|11\rangle$ held by Alice and Bob and possess a singlet-like
structure.
Whenever one party (Alice or Bob) measures the state, the state
decoheres and off-diagonal elements vanish which leads to a
symmetric extendible state \cite{MNPH}:
\begin{equation}
\Upsilon_{AB}=\frac{d}{2d-1}P_{+}+\frac{1}{2d-1}\sum^{d-1}_{i=1}|i\;0\rangle\langle
i\;0|
\end{equation}
from which no entanglement nor secret key can be distilled by
means of 1-LOCC \cite{D1,D2,Lutk3,MNPH}. Therefore, applying
Theorem 2 one derives $K_{\rightarrow}(\Upsilon_{AB})=0$ and
$K_{\rightarrow}(\rho_{AB})\leq
K_{\rightarrow}\downarrow(\rho_{AB})=4$.

\textit{Example 2.} Let us consider a graph state \cite{Hein}
$|\mathcal{G}\rangle$ of a $3n+1$-qubit system associated with a
mathematical graph $\mathcal{G}= \{\mathcal{V},\mathcal{E}\}$,
composed of a set $\mathcal{V}$ of $3n+1$ vertices and a set
$\mathcal{E}$ of edges $\{i,j\}$ connecting each vertex $i$ with
some other $j$:
\begin{equation}
|\mathcal{G}\rangle=\bigotimes_{{i,j}\in\mathcal{E}}CZ_{ij}|\mathcal{G}_{0}\rangle
\end{equation}
where $3n+1$ qubits are initialized in the product state
$|\mathcal{G}_{0}\rangle=\bigotimes_{i\in\mathcal{V}}|\psi_{i}\rangle$
with $|\psi_{i}\rangle=|0_{i}\rangle+|1_{i}\rangle$. Afterwards,
one applies a maximally-entangling control-Z (CZ) gate to all
pairs $\{i,j\}$ of qubits joined by an edge:
$CZ_{ij}=|0_{i}0_{j}\rangle\langle0_{i}0_{j}| +
|0_{i}1_{j}\rangle\langle0_{i}1_{j}|+|1_{i}0_{j}\rangle\langle
1_{i}0_{j}|-|1_{i}1_{j}\rangle\langle1_{i}1_{j}|$. If Alice takes
no more than $n$ qubits from the graph system that will use to
establish communication with Bob who uses other $n$ qubits in this
graph state, then they will be not able by any means to set secure
one-way communication. This results from the fact that the state
$\rho^{AB}_{2n}$ (with n qubits on Alice side and n qubits on
Bob's side) is symmetric extendible to a state $\rho^{AB}_{3n}$
which means that $K_{\rightarrow}(\rho^{AB}_{2n})=0$. A natural
symmetric extension of $\rho^{AB}_{2n}$ is a state
$\rho^{AB}_{3n}=Tr_{B'}|\mathcal{G}\rangle\langle\mathcal{G}|$
resulting from tracing out an arbitrary chosen qubit B' from graph
$\mathcal{G}$. However, if Alice takes $n$ qubits and Bob takes
$n+1$ qubits from the graph system, the resulting state
$\rho^{AB}_{2n+1}$ is not symmetric extendible anymore. Exemplary,
for $n=2$ this state has spectral representation:
\begin{equation}\label{state1}
\rho^{AB}_{2n+1}=\frac{1}{2}(|\phi_{0}\rangle\langle\phi_{0}|+|\phi_{1}\rangle\langle\phi_{1}|)
\end{equation}
where
$|\phi_{0}\rangle=|0_{A}\rangle|0_{B}\rangle+|1_{A}\rangle|1_{B}\rangle$,
$|\phi_{1}\rangle=|0_{A}\rangle|1_{B}\rangle-|1_{A}\rangle|0_{B}\rangle$
and
$\{|0\rangle_{A}=|00-01-10-11\rangle_{A},|1\rangle_{A}=|00+01+10-11\rangle_{A},
|0\rangle_{B}=|001+010+100-111\rangle_{B},|1\rangle_{B}=|000-011-101-110\rangle_{B}\}$.
This state is isomorphic to qubit bipartite state and meets the
condition \cite{Lutk1, Lutk2} for $\cal{C}^{2}\otimes\cal{C}^{2}$
Bell-diagonal states to be symmetric extendible:
$4\sqrt{det(\rho_{AB})} \geq Tr(\rho^{2}_{AB})-\frac{1}{2}$. One
can easily show the isomorphism of $\rho^{AB}_{2n+1}$ for any n
with a qubit bipartite state structure (\ref{state1}). Thus, for
one-way secret key of the state there holds:
$K_{\rightarrow}(\rho^{AB}_{2n+1}) \leq
K_{\rightarrow}\downarrow(\rho^{AB}_{2n+1})=4$, since after
discarding one qubit B' on Bob's side his system would become
symmetric extendible.

\textit{\bf An upper bound on quantum channel capacity.} The best
known definition of the one-way quantum channel capacity
$\mathcal{Q}_{\rightarrow}(\Lambda)$ \cite{Bennett2, Barnum3} is
expressed as an asymptotic regularization of coherent information:
$\mathcal{Q}_{\rightarrow}(\Lambda)=\lim_{n\rightarrow
\infty}\frac{1}{n}\sup_{\rho_{n}} I_{c}(\rho_{n}, \Lambda^{\otimes
n})$ with parallel use of N copies of $\Lambda$ channel. Coherent
information for a channel $\Lambda$ and a source
 state $\sigma$ transferred through the channel is defined as:
$I_{c}(\sigma, \Lambda)=I^{B}(I\otimes \Lambda)(|\Psi\rangle\langle\Psi|)$ where $\Psi$ is a pure state with reduction
$\sigma$ and coherent information of a bipartite state $\rho_{AB}$ shared between Alice and Bob is defined as:
$I^{B}(\rho_{AB})=S(B)-S(AB)$. We will use further the following notation: $I_{c}(A \rangle B)=I^{B}(\rho_{AB})$.

\textbf{Observation 1. }\textit{For a bipartite state
$\rho_{ABB'}\in B(\cal{H}_{A}\otimes \cal{H}_{B}\otimes
\cal{H}_{B'})$ shared between Alice and Bob (B and B' system)
there holds:}
\begin{equation}
I_{c}(A \rangle BB')\leq I_{c}(A \rangle B) + 2S(B')
\end{equation}
\textit{Proof.} One can easily observe that for subadditivity of
entropy $S(BB')\leq S(B)+S(B')$ and for the Araki-Lieb inequality
$|S(AB)-S(B')|\leq S (ABB')$, the left hand side can be bounded as
follows: $S(BB')-S(ABB')\leq S(B)+S(B')-S(AB)+ S(B')=I_{c}(A
\rangle B) + 2S(B')$ which completes the proof. $\Box$

Motivated by the reduced quantity of secret key rate and above
observation we derive further the reduced version of quantum
channel capacity and show that it is a good bound on quantum
channel capacity:

\textbf{Definition 4.} \textit{For a one-way quantum channel
$\Lambda_{BB'}:B(\cal{H}_{BB'})\rightarrow
B(\cal{H}_{\widetilde{B}\widetilde{B'}})$ the reduced one-way
quantum channel capacity is defined as:
\begin{equation}
\mathcal{Q}_{\rightarrow}^{(1)}\downarrow(\Lambda_{BB'}) =
\inf_{\cal{U}}[\mathcal{Q}_{\rightarrow}^{(1)}(\cal{U}(\Lambda_{B}))+
\Delta_{\mathcal{Q}_{\rightarrow}}]
\end{equation}
where $\cal{U}$ denotes unitary operations on Bob's system with a
possible transfer of subsystems from Bob to Eve after action of
$\Lambda_{BB'}$ channel, i.e.
$\cal{U}(\Lambda_{B}(\rho_{B}))=Tr_{B'}\cal{U}\Lambda_{BB'}(\rho_{BB'})$.
$\Delta_{\mathcal{Q}_{\rightarrow}}=2
\sup_{\rho_{BB'}}S(Tr_{B}\Lambda_{BB'}(\rho_{BB'}))$ denotes the
defect parameter related to increase of entropy produced by the
transfer of B'-subsystem from Bob's side to Eve.}

\textbf{Theorem 3. }\textit{For any one-way quantum channel
$\Lambda_{BB'}:B(\cal{H}_{BB'})\rightarrow
B(\cal{H}_{\widetilde{B}\widetilde{B'}})$ there holds:
\begin{equation}
\mathcal{Q}_{\rightarrow}(\Lambda_{BB'}) \leq
\mathcal{Q}_{\rightarrow}\downarrow(\Lambda_{BB'})
\end{equation}
where
$\mathcal{Q}_{\rightarrow}\downarrow(\Lambda_{BB'})=\lim_{n}\mathcal{Q}_{\rightarrow}^{(1)}\downarrow(\Lambda_{BB'}^{\otimes
n })/n$ denotes the reduced quantum capacity. Particularly, for
identity operation $\cal{U}=id$ on Bob's side one obtains:
$\mathcal{Q}_{\rightarrow}(\Lambda_{BB'}) \leq
\mathcal{Q}_{\rightarrow}(\Lambda_{B})+
2\sup_{\rho_{BB'}}S(Tr_{B}\Lambda_{BB'}(\rho_{BB'}))$}.

To prove this inequality for regularized quantum capacity and its
reduced version it is sufficient to derive the below lemma for a
single copy case in analogy to the lemma for one-way secret key
rate above:

\textbf{Lemma 4. }\textit{For any one-way quantum channel
$\Lambda_{BB'}:B(\cal{H}_{BB'})\rightarrow
B(\cal{H}_{\widetilde{B}\widetilde{B'}})$ there holds:
\begin{equation}
\mathcal{Q}_{\rightarrow}^{(1)}(\Lambda_{BB'}) \leq
\mathcal{Q}_{\rightarrow}^{(1)}\downarrow(\Lambda_{BB'})
\end{equation}
}

\textit{Proof.} The proof of this lemma is straightforward with
application of Observation 1 that for a state $\rho_{BB'}$
maximizing coherent information on the left hand side of the
observation the above formula holds also for a possible transfer
of B' to the environment. It is worth recalling that action of
unitary operator on a state does not change its entropy and in a
result coherent information for any partition of the system.$\Box$

Further, one can complete the proof of the theorem in the
asymptotic regime:

\textit{Proof.} To prove the inequality of Theorem 3
asymptotically it suffices to notice that statements of Lemma 4
hold also for arbitrary chosen state $\rho_{n}=\rho^{\otimes n}$.
Now we can prove that: $\mathcal{Q}_{\rightarrow}(\Lambda_{BB'})
\leq \mathcal{Q}_{\rightarrow}(\Lambda_{B})+
\Delta_{\mathcal{Q}_{\rightarrow}}$. Let $\rho^{BB'}_{n}$ be a
state maximizing $\mathcal{Q}_{\rightarrow}(\Lambda_{BB'})$ as an
asymptotic regularization of coherent information, i.e.
$\mathcal{Q}_{\rightarrow}(\Lambda_{BB'})=\lim_{n\rightarrow
\infty}\frac{1}{n}I_{c}(\rho^{BB'}_{n}, \Lambda_{BB'}^{\otimes
n})$ which one can represent as $I_{c}(A \rangle BB')$ for the
aforementioned Choi-Jamiolkowski isomorphism between states and
channels. Basing on Observation 1, one can immediately derive for
the maximizing state $\rho^{BB'}_{n}$: $\frac{1}{n}I_{c}(A \rangle
BB') \leq \frac{1}{n}[I_{c}(A \rangle B) +2S(\rho^{B'}_{n})]$
where $I_{c}(A \rangle
B)=I_{c}(Tr_{B'}\rho^{BB'}_{n},\Lambda^{\otimes n}_{B})$ and
$\rho^{B'}_{n}=Tr_{B}\Lambda_{BB'}^{\otimes n}(\rho^{BB'}_{n})$.
 However, if there exists a state $\sigma^{B}_{n}$ for which $I_{c}(\sigma^{B}_{n},\Lambda^{\otimes n}_{B}) > I_{c}(Tr_{B'}\rho^{BB'}_{n},\Lambda^{\otimes n}_{B})$, then it proves that right hand side of the inequality
in the lemma can be only larger than in case of the chosen state
$\rho^{BB'}_{n}$ which completes the proof. Finally, as in the
aforementioned proof for key subadditivity of entropy can be
applied to verify that in case of the regularized reduced secret
key its defect parameter cannot be larger than
$\Delta_{\mathcal{Q}_{\rightarrow}}=2
\sup_{\rho_{BB'}}S(Tr_{B}\Lambda_{BB'}(\rho_{BB'}))$, since
$\sup_{\rho_{BB'}^{n}}S(Tr_{B^{n}}\Lambda_{BB'}^{\otimes n
}(\rho_{BB'}^{n}))\leq
n\sup_{\rho_{BB'}}S(Tr_{B}\Lambda_{BB'}(\rho_{BB'})$. $\Box$

\textit{Example 3.} As an example we will use the aforementioned
graph state from Example. 2 and we will search for one-way channel
capacity of a channel $\Lambda_{BB'}$, isomorphic due to
Choi-Jamiolkowski isomorphism, with a state
$\rho^{ABB'}_{2n+1}=(I\otimes
\Lambda_{BB'})|\Psi\rangle\langle\Psi|$. As above, after
discarding $B'$ 1-qubit system the state would become symmetric
extendible that implies $Q_{\rightarrow}(\Lambda_{B})=0$.
Therefore, we obtain $Q_{\rightarrow}(\Lambda_{BB'})\leq 2$.

The power of the above results appears especially in application
of Lemma 3 to any channel reducible to anti-degradable channel
which Choi-Jamiolkowski representation is symmetric extendible
\cite{Lutk1} or channels reducible to degradable channels which
have known capacity \cite{Smith1}.

\textit{\bf Dual picture for one-way distillable entanglement and
private information.} Our results for one-way secret key and
quantum channel capacity lead immediately to similar reduced
formula for private information and one-way distillation
quantities. The private capacity \cite{DevetakW3, DevetakW4}
$\mathcal{P}(\Lambda)$ of a quantum channel is equal to
regularization of private information:
$\mathcal{P}^{(1)}(\Lambda)=\max_{X,\rho_{x}^{A}}(I(X,B)-I(X,E))$
with maximization over classical random variables X and input
quantum states $\rho_{x}^{A}$ depending on the value of X.
Absorbing T into X variable in Theorem 1. leads to definitions for
private information and private capacity \cite{DevetakW4}, thus,
following Lemma 3. we can derive an upper bound on private
information and private capacity via their reduced counterparts:

\textbf{Definition 5.} \textit{For a one-way quantum channel
$\Lambda_{BB'}:B(\cal{H}_{BB'})\rightarrow
B(\cal{H}_{\widetilde{B}\widetilde{B'}})$ the reduced private
information is defined as:
\begin{equation}
\mathcal{P}^{(1)}\downarrow(\Lambda_{BB'}) =
\inf_{\cal{U}}[\mathcal{P}^{(1)}(\cal{U}(\Lambda_{B}))+
\Delta_{P}]
\end{equation}
where $\cal{U}$ denotes unitary operations on Bob's system with a
possible transfer of subsystems from Bob to Eve, i.e.
$\cal{U}(\Lambda_{B}(\rho_{B}))=Tr_{B'}\cal{U}\Lambda_{BB'}(\rho_{BB'})$.
$\Delta_{P} =4 S (\rho_{B'})$ denotes the defect parameter related
to increase of entropy produced by the transfer of B'-subsystem
from Bob's side to Eve.}

\textbf{Theorem 4. }\textit{For a one-way quantum channel
$\Lambda_{BB'}:B(\cal{H}_{BB'})\rightarrow
B(\cal{H}_{\widetilde{B}\widetilde{B'}})$ there holds:
\begin{equation}
\mathcal{P}(\Lambda_{BB'}) \leq
\mathcal{P}\downarrow(\Lambda_{BB'})
\end{equation}
where
$\mathcal{P}\downarrow(\Lambda_{BB'})=\lim_{n}\mathcal{P}^{(1)}\downarrow(\Lambda_{BB'}^{\otimes
n })/n$ denotes the reduced private capacity. Particularly, for
identity operation $\cal{U}=id$ on Bob's side one obtains:
$\mathcal{P}(\Lambda_{BB'}) \leq \mathcal{P}(\Lambda_{B})+
4S(\rho_{B'})$}

The proof can be conducted in analogy to Theorem 2. and Lemma 3,
however, for regularization of reduced private information it is
crucial to derive the below lemma for a one-copy case:

\textbf{Lemma 5. }\textit{For every one-way quantum channel
$\Lambda_{BB'}:B(\cal{H}_{BB'})\rightarrow
B(\cal{H}_{\widetilde{B}\widetilde{B'}})$ there holds:
\begin{equation}
\mathcal{P}^{(1)}(\Lambda_{BB'}) \leq
\mathcal{P}^{(1)}\downarrow(\Lambda_{BB'})
\end{equation}
}

\textit{Proof.} To prove this lemma it suffices to absorb variable
T into X in Theorem 1. for definition of private information and
conduct the proof in analogy to the proof of Lemma 2 for a channel
$\Lambda_{BB'}$ and a chosen state $\rho$ sent through it. $\Box$

We can now propose a new bound on distillation of entanglement by
means of one-way LOCC. This result is based on observation
\cite{DevetakW3, DevetakW4} that one-way distillable entanglement
$D_{\rightarrow}$ of a state $\rho_{AB}$ can be represented as
regularization of one-copy formula:
$D^{(1)}_{\rightarrow}(\rho_{AB})=\max_{\textbf{T}}\sum_{l=1}^{L}\lambda_{l}I_{c}(A\rangle
B)_{\rho_{l}}$ where the maximization is over quantum instruments
$T = (T_{1}, \dots , T_{L})$ on Alice's system,
$\lambda_{l}=TrT_{l}(\rho_{A})$, $T_{l}$ is assumed to have one
Kraus operator $T_{l}(\rho)=A_{l}\rho A_{l}^{\dag}$ and
$\rho_{l}=\frac{1}{\lambda_{l}}(T_{l}\otimes id)\rho_{AB}$. Basing
on the results of Observation 1. and Lemma 3. we derive a general
formula for the bound on one-way distillable entanglement applying
the reduced quantity:

\textbf{Definition 4.} \textit{For a bipartite state
$\rho_{ABB'}\in B(\cal{H}_{A}\otimes \cal{H}_{B}\otimes
\cal{H}_{B'})$ shared between Alice and Bob (B and B' system) the
reduced one-way distillable entanglement is defined as:
\begin{equation}
D_{\rightarrow}^{(1)}\downarrow(\rho_{ABB'}) =
\inf_{\cal{U}}[D_{\rightarrow}^{(1)}(\cal{U}(\rho_{AB}))+
\Delta_{D_{\rightarrow}}]
\end{equation}
where $\cal{U}$ denotes unitary operations on Bob's system with a
possible transfer of subsystems from Bob to Eve, i.e.
$\cal{U}(\rho_{AB})=Tr_{B'}(I\otimes \cal{U})\rho_{ABB'}$.
$\Delta_{D_{\rightarrow}} =2 S (\rho_{B'})$ denotes the defect
parameter related to increase of entropy produced by the transfer
of B'-subsystem from Bob's side to Eve.}

\textbf{Theorem 5. }\textit{For a bipartite state $\rho_{ABB'}\in
B(\cal{H}_{A}\otimes \cal{H}_{B}\otimes \cal{H}_{B'})$ shared
between Alice and Bob (B and B' system) there holds:
\[
D_{\rightarrow}(\rho_{ABB'}) \leq
D_{\rightarrow}\downarrow(\rho_{ABB'})
\]
where $\Delta_{D_{\rightarrow}}=2S(\rho_{B'})$ and
$D_{\rightarrow}\downarrow(\rho_{ABB'})=\lim_{n}D_{\rightarrow}^{(1)}\downarrow(\rho_{ABB'}^{\otimes
n })/n$ denotes regularized version of reduced one-way distillable
entanglement for one copy. Particularly, for identity operation
$\cal{U}=id$ on Bob's side one obtains:
$D_{\rightarrow}(\rho_{ABB'}) \leq D_{\rightarrow}(\rho_{AB})+
2S(\rho_{B'})$.}

The proof of this theorem can be conducted in analogy to the
previous proofs for bounds on one-way secret key and quantum
channel capacity. The left inequality is an immediate implication
of the following lemma for the one-copy formula:

\textbf{Lemma 6. }\textit{For every bipartite state $\rho_{ABB'}$
there holds:
\begin{equation}
D_{\rightarrow}^{(1)}(\rho_{ABB'}) \leq
D_{\rightarrow}^{(1)}\downarrow(\rho_{ABB'})
\end{equation}
}
\textit{Proof.} It suffices to use results of Observation 1. to
notice that for a chosen set of instruments $\textbf{T}$ on Alice
side for calculation of $D_{\rightarrow}^{(1)}(\rho_{ABB'})$ the
inequality holds as extension of inequality from Observation 1. by
multiplicands $\lambda_{l}$ on the left and right side. However,
if in case of calculating $D_{\rightarrow}^{(1)}(\rho_{AB})$ there
exists a set $\textbf{T'}$ maximizing $D_{\rightarrow}(\rho_{AB})$
better than $\textbf{T}$, then right hand side of the inequality
can be only greater. $\Box$

It is crucial to notice that the 'defect' parameters $\Delta$ for
the reduced quantities are subadditive and hence, can be exploited
in case of composite systems and regularization:

\textit{\bf Corollary. }\textit{For the reduced quantities of
$\{K_{\rightarrow},\mathcal{P}, \mathcal{Q}_{\rightarrow},
D_{\rightarrow}\}$ for composite systems there holds:
$\Delta_{X}(\rho\otimes\sigma)\leq\Delta_{X}(\rho)+\Delta_{X}(\sigma)$
and
$\Delta_{Y}(\Lambda\otimes\Gamma)\leq\Delta_{Y}(\Lambda)+\Delta_{Y}(\Gamma)$
where $X=\{K_{\rightarrow},D_{\rightarrow}\}$ stands for states
and $Y=\{\mathcal{Q}_{\rightarrow},\mathcal{P}\}$ for channels
respectively. }

To prove the above corollary it suffices to use subadditivity of
entropy for composite systems since Bob can act with a unitary
operation before he discard some part of his subsystem. This
property of the parameters enables regularization in the
asymptotic regime of the reduced quantities for large systems
$\rho^{\otimes n}$.

\textit{Example 4. Activable multi-qubit bound entangled states.}
As an example illustrating this bound we consider an activated
bound entangled state $\rho_{II}$ \cite{Dur} which is distillable
if the parties (Alice and Bob) form two groups containing between
$40\%$ and $60\%$ of all parties of the system in the state
$\rho_{II}$. If Alice or Bob posses less than $40\%$ of the system
or system is shared between more than two parties, then the state
becomes undistillable. This state for large amount of particles
can manifest features characteristic for 'macroscopic
entanglement' with no 'microscopic entanglement'. For definition
of the state, let us consider the family $\rho_{N}$ of N-qubit
states:
$\rho=\sum_{\sigma=\pm}\lambda_{0}^{\sigma}|\Psi_{0}^{\sigma}\rangle\langle\Psi_{0}^{\sigma}|+
\sum_{k\neq0}\lambda_{k}(|\Psi_{k}^{+}\rangle\langle\Psi_{k}^{+}|+|\Psi_{k}^{-}\rangle\langle\Psi_{k}^{-}|)$
where $|\Psi_{k}^{\pm}\rangle=\frac{1}{\sqrt{2}}(|k_{1}k_{2}\ldots
k_{N-1}0\rangle\pm|\overline{k}_{1}\overline{k}_{2}\ldots
\overline{k}_{N-1}1\rangle)$ are GHZ-like states with
$k=k_{1}k_{2}\ldots k_{N-1}$ being a chain of $N-1$ bits and
$k_{i}=0, 1$ if $\overline{k}_{i}=1, 0$, thus, the state is
parameterized by $2^{N-1}$ coefficients. Let us consider now a
bipartite splitting $\mathcal{P}$ where Alice takes $0.6N$ of
qubits and Bob takes the other $0.4 N$ qubits. We can immediately
show that $D_{\rightarrow}(\rho_{II})\leq -
2(\lambda_{0}^{\pm}+2\sum_{k}\lambda_{k})\log(\lambda_{0}^{\pm}+2\sum_{k}\lambda_{k})$
since for Bob transferring one qubit to the environment, we obtain
undistillable state $D_{\leftrightarrow}(\rho_{N-1})=0$. It is
noticeable that even for a large macroscopic system with
$N\rightarrow\infty$, $D_{\rightarrow}(\rho_{II})\leq - 2
(\lambda_{0}^{\pm}+2\sum_{k}\lambda_{k})\log(\lambda_{0}^{\pm}+2\sum_{k}\lambda_{k})$.
It can be easily shown that with the same method it is possible to
achieve an upper bound on one-way quantum channel capacity
$Q_{\rightarrow}$.

\textit{\bf Conclusions.} In this paper we proposed new reduced
versions of quantum quantities: reduced one-way quantum key,
distillable entanglement and reduced corresponding capacities. We
show that in some cases they may provide bounds on the non-reduced
versions simplifying drastically their estimations. It is evident
especially in case of states of large systems which is supported
by examples. The open problem is whether they can be applied to
non-additivity problem of quantum channel capacities and quantum
secure key \cite{Smith, Smith1}. Further, it is not known if they
have analogs in general quantum networks and whether the bounds
can be improved by better estimation of defect parameters.

\textit{\bf Acknowledgments.} The authors thank Michal Horodecki
for critical comments on this paper. This work was supported by
Ministry of Science and Higher Education grant No N202 231937.
Part of this work was done in National Quantum Information Center
of Gdansk.


\begin{thebibliography}{8}
\bibitem{BennettDiVincenzo}
C. H. Bennett, D. P. DiVincenzo, J. A. Smolin and W. K. Wootters
Phys. Rev. A 5, 3824 (1996)
\bibitem{BennettSmolin}
C. H. Bennett, G. Brassard, S. Popescu, B. Schumacher, J. A.
Smolin and W. K. Wootters, Phys. Rev. Lett. 76, 722 (1996).
\bibitem{Barnum3}
H.Barnum, E. Knill and M. A. Nielsen, IEEE Trans. Info. Theor. 46
1317 (2000).
\bibitem{Barnum4}
H. Barnum, M. Nielsen and B. Schumacher Phys. Rev. A {\bf 57},
4153 (1998).
\bibitem{DevetakW1}
I. Devetak, A. Winter, Proc. R. Soc. Lond. A  {\bf 461}, 207
(2005).
\bibitem{DevetakW2}
I. Devetak, A. Winter, Phys. Rev. Lett. {\bf 93}, 080501 (2004).
\bibitem{DevetakW3}
I. Devetak, IEEE Trans. Inform. Theory {\bf 51}, 44 (2005).
\bibitem{KHPH}
K. Horodecki et al., Phys. Rev. Lett. {\bf 94}, 200501 (2005).
\bibitem{KHPH2}
K. Horodecki et al., IEEE Trans. Inf. Theory {\bf 55}, 1898
(2009).
\bibitem{Renner}
R. Renner and S. Wolf, Advances in Cryptology - EUROCRYPT '03,
Lecture Notes in Computer Science {\bf 2656}, 562 (2003).
\bibitem{Smith} G. Smith and J. Yard, Science {\bf 321}, 1812 (2008).
\bibitem{DiVincenzo}
D. P. DiVincenzo et al., Phys. Rev. Lett. {\bf 92}, 067902 (2004).

\bibitem{Entropy}
For any quantum state $\rho$ one can define a concave function $S(\rho)\equiv -Tr(\rho\log\rho)$ called
as the von Neumann entropy and its classical counterpart Shannon entropy for a probability distribution P:
$H(P)\equiv - \sum_{x}P(x)\log P(x)$.
\bibitem{MInformation}
For any bipartite state $\rho_{AB}$ one defines the quantum mutual information: $I(A:B)=S(A)+S(B)-S(AB)$
and further, for a tripartite system $\rho_{ABC}$ the conditional quantum mutual information: $I(A:B|C)=S(AC)+S(BC)-S(ABC)-S(C)$ where we use the notation for entropy of X system $S(\rho_{X})=S(X)$.
\bibitem{HolevoFunction}
The Holevo function $\chi(\cdot)$ is defined for any ensemble of density matrices
$\mathfrak{A}=\{p_{i}, \rho_{i}\}$ with average density matrix $\rho=\sum_{i}p_{i}\rho_{i}$ as follows:
$\chi(\rho)=S(\sum_{i}p_{i}\rho_{i})-\sum_{i}p_{i}S(\rho_{i})$
and is a good upper bound \cite{Holevo1, Holevo2} on the accessible information.
\bibitem{Ekert}
M. Christland, A. Ekert et al., Proceedings of the 4th Theory of
Cryptography Conference, Lecture Notes in Computer Science {\bf
4392}, 456 (2007).
\bibitem{Bruss}
D. Bruss, D. P. DiVincenzo, A. Ekert, C. A. Fuchs, C. Macchiavello
and J. A. Smolin, Phys. Rev. A {\bf 57}, 2368 (1998).
\bibitem{MNPH}
M. L. Nowakowski, P. Horodecki, J. Phys. A {\bf 42}, 135306
(2009).
\bibitem{MNPH2}
M. L. Nowakowski, In preparation.
\bibitem{Lutk3}
T. Moroder, N. L\"{u}tkenhaus, Phys. Rev. A {\bf 74}, 052301
(2006).

\bibitem{D1}
A. C. Doherty, P. A. Parillo and F. M. Spedalieri, Phys. Rev.
Lett. {\bf 88}, 187904 (2002).
\bibitem{D2}
A. C. Doherty, P. A. Parillo and F. M. Spedalieri, Phys. Rev. A
{\bf 69}, 022308 (2004).
\bibitem{Hein}
M. Hein et al., Entanglement in Graph States and Its Applcations,
Preprint quant-ph/0602096.
\bibitem{Lutk1}
G. O. Myhr, N. L\"{u}tkenhaus, Spectrum conditions for symmetrc
extendible states, Preprint quant-ph/0812.3667v1.
\bibitem{Smith1}
G. Smith, J. A. Smolin, A. Winter, IEEE Trans. Info. Theory {\bf
54 }, 4208 (2008).
\bibitem{Lutk2}
G. O. Myhr et. al., Symmetric extensions in two-way quantum key
distribution, Preprint quant-ph/0812.3607v1.
\bibitem{Bennett2} C. H. Bennett, D. P. Di Vincenzo, J. Smolin and W. K.
Wootters, Phys. Rev. A {\bf 54}, 3814 (1997).

\bibitem{DevetakW4}
I. Devetak, The private classical capacity and quantum capacity of
a quantum channel, Preprint quant-ph/9809023.


\bibitem{Dur}
W. D\"{u}r, J. I. Cirac, Phys. Rev. A {\bf 62}, 022302 (2000).

\bibitem{Holevo1}
A. S. Holevo, Coding theorems for quantum channels, Preprint quant-ph/9809023.
\bibitem{Holevo2}
A. S. Holevo, Probl. Inf. Transm. USRR {\bf 9}, 31-42 (1973).








\end{thebibliography}
\end{document}